\documentclass[10pt,preprint]{emulateapj}
\usepackage{natbib}
\def\apj{ApJ}
\def\aj{AJ}

\def\nat{Nature}

\def\pasp{PASP}

\newcommand{\arcm}{\ifmmode{'}\else$'$\fi}
\newcommand{\arcs}{\ifmmode{''}\else$''$\fi}

\shorttitle{No evidence for Cepheids and dwarf galaxy behind the Galactic disk}
\shortauthors{Pietrukowicz et al.}

\begin{document}

\title{No Evidence for Classical Cepheids and a New Dwarf Galaxy \\
Behind the Galactic Disk\altaffilmark{1}}

\author{P.~Pietrukowicz\altaffilmark{2},
A.~Udalski\altaffilmark{2},
M.~K.~Szyma\'nski\altaffilmark{2},
I.~Soszy\'nski\altaffilmark{2},
G.~Pietrzy\'nski\altaffilmark{2,3},
{\L}.~Wyrzykowski\altaffilmark{2},
R.~Poleski\altaffilmark{2,4},
K.~Ulaczyk\altaffilmark{2,5},
J.~Skowron\altaffilmark{2},
P.~Mr\'oz\altaffilmark{2},
M.~Pawlak\altaffilmark{2},
and S.~Koz{\l}owski\altaffilmark{2}
}
\altaffiltext{1}
{Based on observations obtained with the 1.3-m Warsaw telescope
at the Las Campanas Observatory of the Carnegie Institution for Science}
\altaffiltext{2}
{Warsaw University Observatory, Al. Ujazdowskie 4, 00-478 Warszawa, Poland}
\altaffiltext{3}
{Universidad de Concepci{\'o}n, Departamento de Astronom\'ia, Casilla 160-C, Concepci{\'o}n, Chile}
\altaffiltext{4}
{Department of Astronomy, Ohio State University, 140 W. 18th Ave., Columbus, OH 43210, USA}
\altaffiltext{5}
{Department of Physics, University of Warwick, Coventry CV4 7AL, UK}

\begin{abstract}

Based on data from the ongoing OGLE Galaxy Variability Survey
(OGLE GVS) we have verified observed properties of stars detected
by the near-infrared VVV survey in a direction near the Galactic
plane at longitude $l\approx-27\degr$ and recently tentatively
classified as classical Cepheids belonging to a, hence claimed,
dwarf galaxy at a distance of about 90~kpc from the Galactic Center.
Three of four stars are detected in the OGLE GVS $I$-band images.
We show that two of the objects are not variable at all and the third
one with a period of 5.695~d and a nearly sinusoidal light curve
of an amplitude of 0.5~mag cannot be a classical Cepheid and is
very likely a spotted object. These results together with a very
unusual shape of the $K_s$-band light curve of the fourth star
indicate that very likely none of them is a Cepheid and, thus,
there is no evidence for a background dwarf galaxy. Our observations
show that a great care must be taken when classifying objects
by their low-amplitude close-to-sinusoidal near-infrared
light curves, especially with a small number of measurements.
We also provide a sample of high-amplitude spotted stars with
periods of a few days that can mimick pulsations and even eclipses.

\end{abstract}


%%%%%%%%%%%%%%%%%%%%%%%%%%%%%%%%%%%%%%%%%%%%%%%%%%%%%%%%%%%%%%%%%%%%%%%%%%%%%%%

\section{The Target Objects}

Recently \cite{2015ApJ...802L...4C} reported on the detection of four classical
Cepheids clustered in angle (within one degree) and distance (on average 90~kpc
from the Galactic Center) in a direction near the Galactic plane
at longitude $l\approx-27\degr$. They used near-infrared data from
the ESO public survey VISTA Variables in the Via Lactea
\citep[VVV,][]{2010NewA...15..433M}. Their search for periodic variables
was based on time-series photometry in the $K_s$ band with only 32
epochs per star across the VVV disk area. For each of the star classified
by them as a Cepheid they provide single epoch VVV photometry in the $JHK_s$
bands and extinction values derived from the color excess assuming the
\cite{1989ApJ...345..245C} extinction law. Finally, they calculate distances
to the objects using a formula given in \cite{2014Natur.509..342F}.
The obtained mean distance of 90 kpc and clumped location in the sky led them
to a conclusion that the stars are associated with a previously
unknown dwarf galaxy. In our work, we show that to conclude on the
presence of such a galaxy one has to be absolutely sure that the observed
stars are of particular type and are indeed located at a similar distance
behind the Galactic disk.

%%%%%%%%%%%%%%%%%%%%%%%%%%%%%%%%%%%%%%%%%%%%%%%%%%%%%%%%%%%%%%%%%%%%%%%%%%%%%%%

\section{Analysis of the OGLE Data}

The OGLE project in its fourth phase of operation \citep[OGLE-IV,][]{2015AcA....65....1U}
conducts, among others, a survey of the Galactic disk area visible from
Las Campanas Observatory, Chile---the OGLE Galaxy Variability Survey (OGLE GVS).
It covers 2/3 of the Galactic plane with longitudes from $-170\degr$
to $+60\degr$ and latitudes $-3\degr \lesssim b \lesssim +3\degr$. Regular
monitoring with a cadence of 1--2 days, called the shallow OGLE GVS,
started in 2013. It spans a magnitude range $10<I<19$. A complementary,
deeper survey started in 2010 and maps the Galactic plane area down
to $I\approx22$~mag. The data will be extremely useful for studying
the Milky Way structure and properties of many variable objects
\citep[cf.][]{2013AcA....63..115P,2013AcA....63..379P}. However, at the moment
of writing (June 2015) sufficient amount of data have been collected
to find information on selected Galactic plane objects,
such as the candidate Cepheids reported in \cite{2015ApJ...802L...4C}.

For clarity, it is very important to note that in their paper,
\cite{2015ApJ...802L...4C} made the following errors in IDs,
coordinates, and finding charts for reported objects, later corrected
in private communication with us: (1) everywhere in their paper the correct ID
for the second reported star should be VVV J162231.35-512346.9 (not
VVV J162328.18-513230.4) and thus the Galactic coordinates given
in their Table~1 are also incorrect; (2) the finding chart for the second
object is wrong, while the $K_s$-band light curve shown in their Fig.~1
is correct; (3) in the same figure, they labeled the fourth object
using the ID of the third one. Our Table~\ref{tab:objects} gives correct
positions of the investigated objects. Among minor things
presented in \cite{2015ApJ...802L...4C} in a different than common manner
are the finding charts in Galactic coordinates instead of equatorial
coordinates and an opposite direction of increasing Galactic longitude
in their Fig.~1.

\begin{figure}[htb!]
\centering
\includegraphics[width=8.5cm]{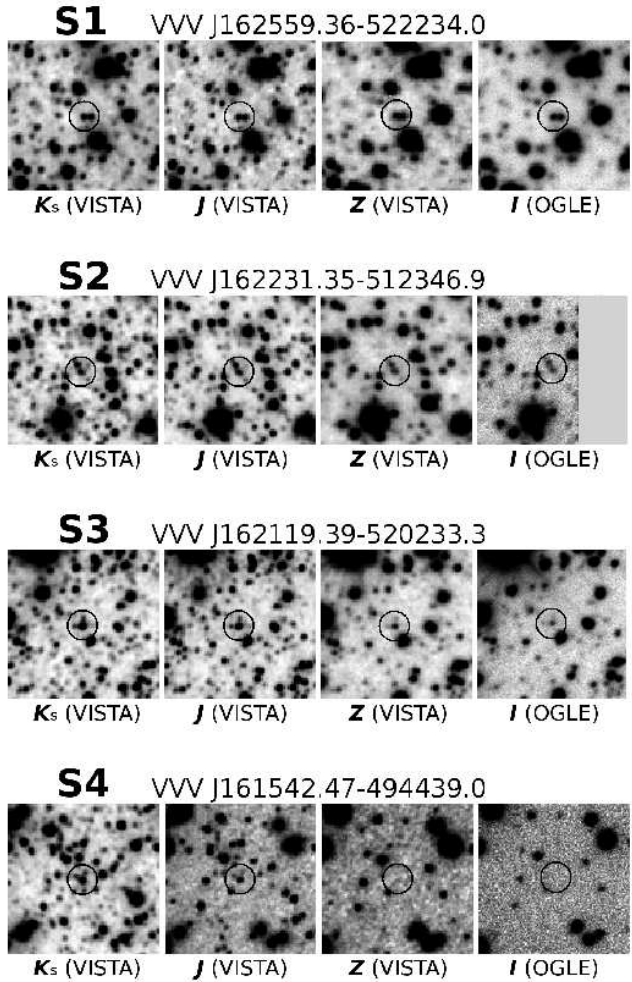}
\caption{Images showing the four target stars (in the centers of the circles)
in four different bands: $K_sJZ$ from the 4.1-m VISTA telescope and in $I$
from the 1.3-m OGLE telescope. Each image is $30\arcs\times30\arcs$.
North is up and East to the left. Only the last target is not visible
in the OGLE GVS data.}
\label{fig:charts}
\medskip
\end{figure}

\begin{figure}[htb!]
\centering
\includegraphics[width=8.5cm]{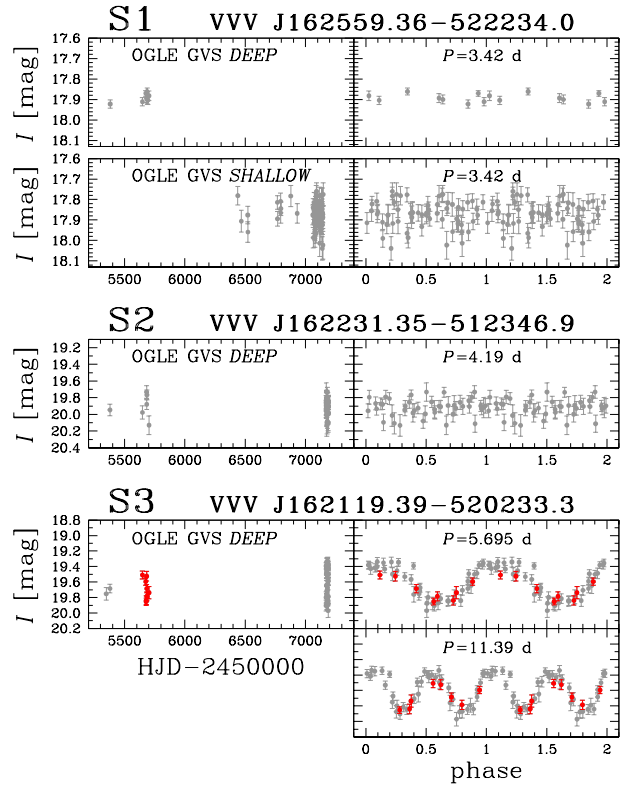}
\caption{Time-domain (left panels) and phased (right panels) $I$-band light
curves of the three stars detected in the OGLE Galaxy Variability Survey.
Objects S1 and S2 clearly do not phase with the periods given in
\cite{2015ApJ...802L...4C}. Deep OGLE observations confirm light
variations in star S3 with a period of 5.695~d (or twice longer),
but their shape is different than expected for fundamental-mode classical
Cepheids. Observations from 2011 for this object are marked in red
to show a likely amplitude change.}
\label{fig:curves}
\medskip
\end{figure}

\begin{figure}[htb!]
\centering
\includegraphics[width=8.5cm]{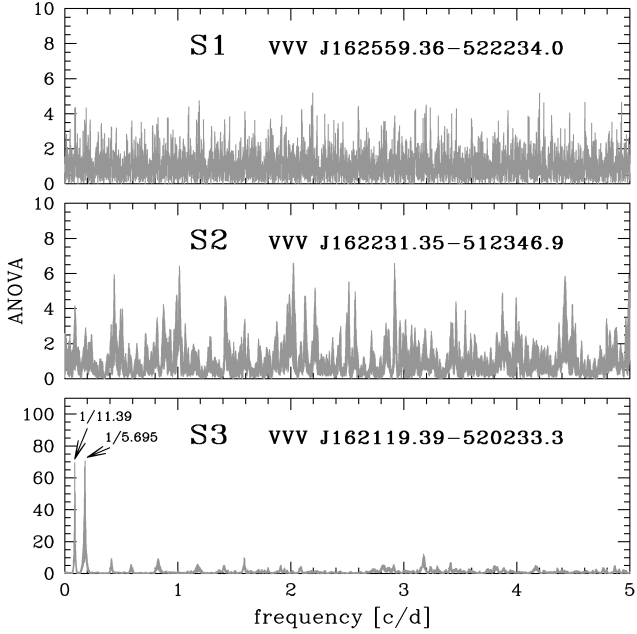}
\caption{ANOVA power spectra of the three objects observed by OGLE GVS.
No periodic signal is detected in stars S1 and S2. Object S3 shows
strong peaks at frequencies $1/5.695$~c/d and $1/11.39$~c/d.}
\label{fig:power}
\medskip
\end{figure}

\begin{table*}[t!]
\centering \caption{Data on individual target stars}
\begin{tabular}{ccccccc}
\hline
Star &       VVV ID        &    $l$    &    $b$    & OGLE field & OGLE ID & $P_{\rm OGLE}$ \\
     &                     &  [\degr]  &  [\degr]  &            &         &    [d]    \\
\hline
S1   & J162559.36-522234.0 & $-27.597$ &  $-2.237$ & GD1126.25  &   6394  &    no     \\
S2   & J162231.35-512346.9 & $-27.273$ &  $-1.168$ & GD1133.28  &  15388  &    no     \\
S3   & J162119.39-520233.3 & $-27.862$ &  $-1.494$ & GD1133.13  &   8832  &  5.6950(3) or $\times2$ \\
S4   & J161542.47-494439.0 & $-26.888$ &  $+0.768$ & GD1139.19  &    -    &    -     \\
\hline
\end{tabular}
\label{tab:objects}
\medskip
\end{table*}

\begin{figure}[htb!]
\centering
\includegraphics[width=8.5cm]{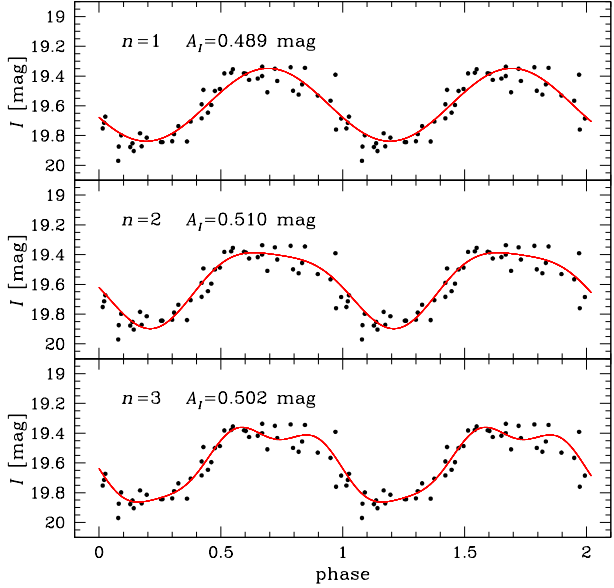}
\caption{Fourier series with increasing order from $n=1$ to $n=3$ fitted
to the $I$-band light curve of star S3 folded with the period $P=5.695$~d.
Full amplitudes $A$ are given to each fit. The fit with $n=2$ seems to be optimal.}
\label{fig:fourier}
\medskip
\end{figure}

\begin{figure}[htb!]
\centering
\includegraphics[width=8.5cm]{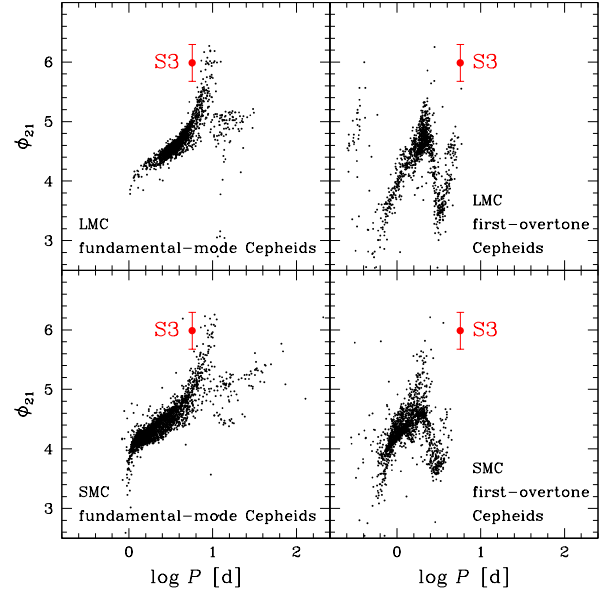}
\caption{Fourier combination $\phi_{21}$ vs. ${\rm log}~P$ for fundamental-mode
(left panels) and first-overtone Cepheids (right panels) from the LMC
(upper panels) and SMC (lower panels). Star S3 with a nearly sinusoidal light
curve cannot be a fundamental-mode pulsator.}
\label{fig:phi21}
\medskip
\end{figure}

\begin{figure}[htb!]
\centering
\includegraphics[width=8.5cm]{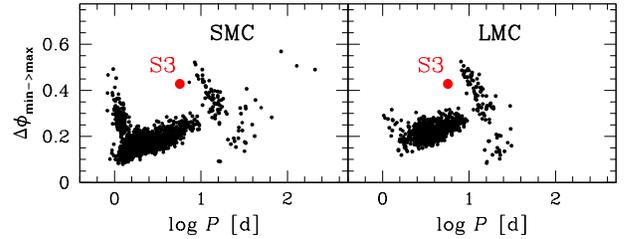}
\caption{Phase difference between minimum and maximum brightness
for fundamental-mode classical Cepheids with $I$-band amplitudes
$>0.4$~mag from SMC (left panel) and LMC (right panel). The observed
abrupt phase difference at about 10~d is related to the effect of
Hertzsprung progression. Star S3 with $P=5.695$~d and nearly
sinusoidal 0.5-mag light curve is a clear outlier.}
\label{fig:steepness}
\medskip
\end{figure}

\begin{figure}[htb!]
\centering
\includegraphics[width=8.5cm]{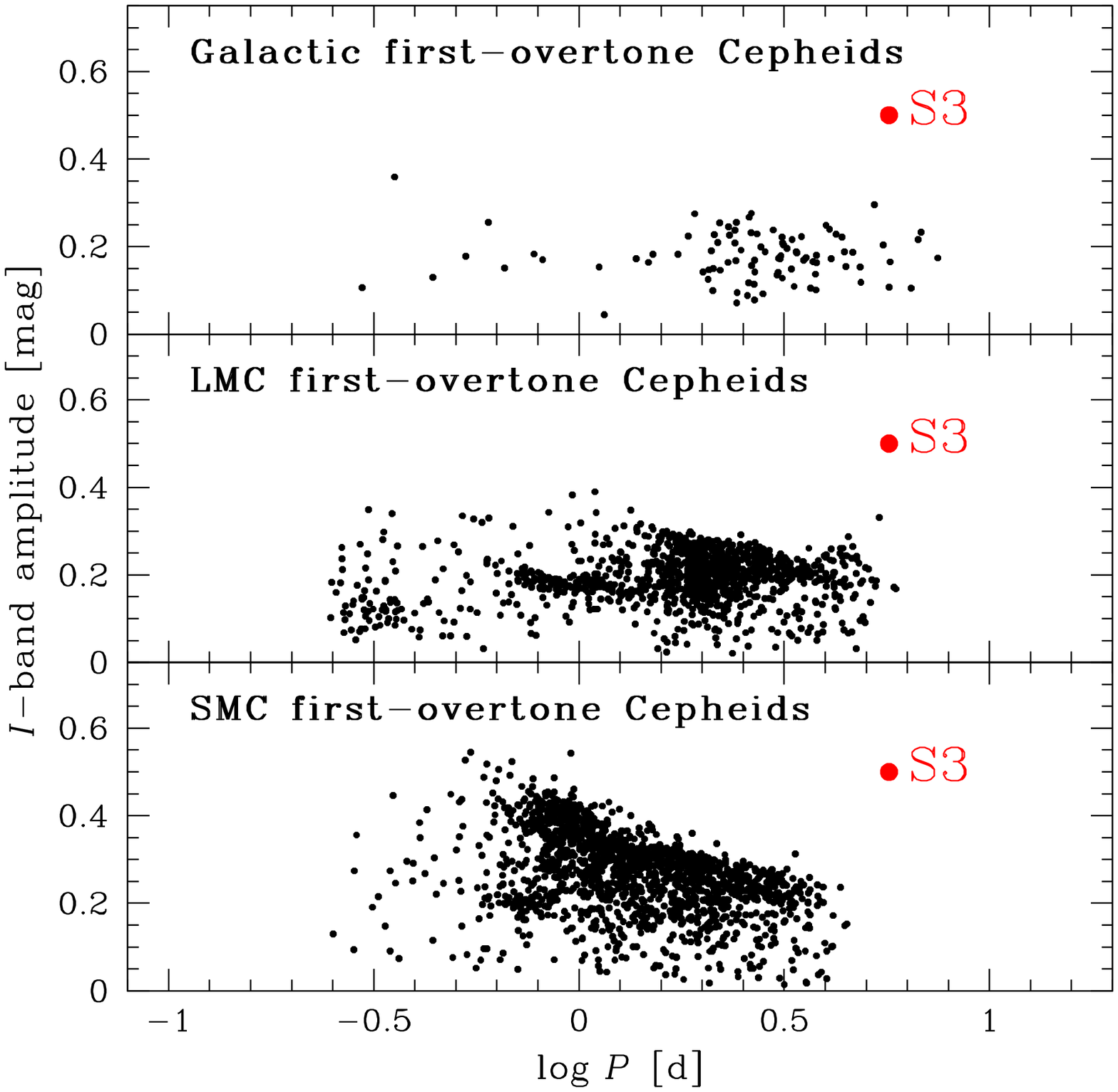}
\caption{Period-amplitude diagram for known first-overtone Cepheids
from three different environments (from top to bottom):
of our Galaxy \citep[ASAS data,][]{2002AcA....52..397P,2014AcA....64..115S},
LMC \citep[OGLE data,][]{2008AcA....58..163S}, and SMC
\citep[OGLE data,][]{2010AcA....60...17S}. Star S3 with the period $P=5.695$~d
has evidently too high amplitude to be such a type of pulsator.}
\label{fig:pamp}
\medskip
\end{figure}

\begin{figure}[htb!]
\centering
\includegraphics[width=8.5cm]{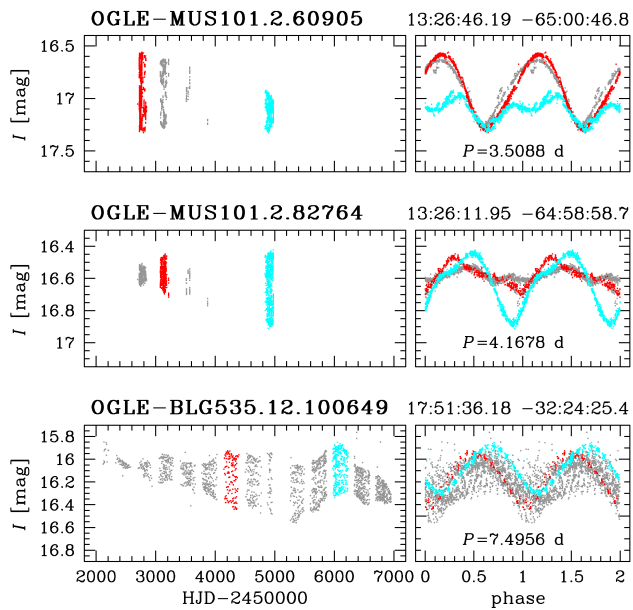}
\caption{Examples of $I$-band light curves of three spotted stars from OGLE.
Identification names, equatorial coordinates, and periods are given 
for each object. Two selected seasons are shown with different colors.
Light curves of spotted stars may have amplitudes exceeding 0.5~mag
and may mimick pulsating stars and even eclipsing binaries if observed
for a short period of time.}
\label{fig:spotted}
\medskip
\end{figure}

\begin{figure}[htb!]
\centering
\includegraphics[width=8.5cm]{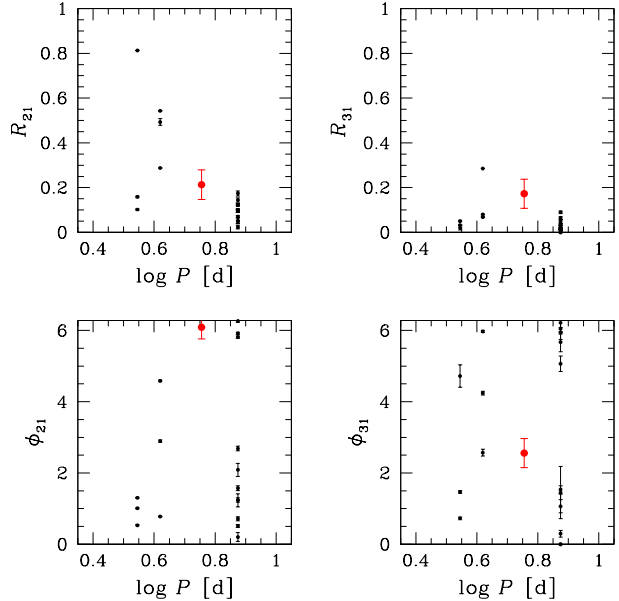}
\caption{Fourier combinations $R_{21}$, $R_{31}$, $\phi_{21}$,
and $\phi_{31}$ in function of period for star S3 (red points) in comparison
to the three spotted stars presented in Fig.~\ref{fig:spotted}. Black
points represent separate seasons of observations. The parameteres describing
the light curve of S3 are in agreement with spotted objects.}
\label{fig:S3spotted}
\medskip
\end{figure}

Once we made sure of the positions of the reported objects,
we found three of the four stars, hereafter denoted as S1, S2, S3,
and S4, in deep OGLE GVS $I$-band frames (see Fig.~\ref{fig:charts}).
Only object S4 located about $0\fdg77$ from the Galactic plane
in a highly reddened area is not visible. Star S1 is bright
enough to be verified for variability in the shallow OGLE GVS images,
but stars S2 and S3 required additional deep observations taken
in May--June 2015. The number of $I$-band measurements for S1, S2,
and S3 is 61, 48, and 44, respectively. The photometry was extracted
using the Difference Image Analysis (DIA) method with an image subtraction
algorithm implemented by \cite{2000AcA....50..421W} to the OGLE images
of dense stellar fields. \cite{2015AcA....65....1U} gives full details
on the data reduction techniques. In Fig.~\ref{fig:curves}, we show
light curves of the three detected stars, while in Fig.~\ref{fig:power},
we present power spectra obtained with the ANOVA statistics
\citep{1996ApJ...460L.107S}. Stars S1 and S2 look to be constant.
There is no significant period in their power spectra. Light curves
clearly do not phase with the periods given in \cite{2015ApJ...802L...4C}.
If the two objects were indeed Cepheids with pulsation periods of 3--4~d
as proposed on the base of the $K_s$-band observations, we would see
them in the $I$ band as large-amplitude variables ($A_I\gtrsim0.2$~mag).
Objects S1 and S2 either are constant stars or have amplitudes
of at most a few hundreds of magnitude.

Deep OGLE GVS observations confirm the variable nature of object S3
with a period around 5.69~d noted in \cite{2015ApJ...802L...4C}.
The power spectrum based on the optical data shows a very strong
signal at $P=5.695$~d and also $2P=11.390$~d. The shape of the light
curve is close to a sinusoid, but its stability is under question.
It is likely that the $I$-band amplitude changed from about 0.35~mag in 2011
to 0.50~mag in 2015. The value for 2011 is, however, uncertain, since
only eight images were taken that year. In Fig.~\ref{fig:fourier},
we show that the optimal Fourier fit to the $I$-band data is of an maximal
order of $n=2$, assuming $P=5.695$~d and stability of the light curve.
For $n=2$ we obtain the following Fourier parameters:
$A_0=19.593\pm0.009$, $A_1=0.251\pm0.016$, $A_2=0.055\pm0.016$,
$\phi_1=5.03\pm0.06$, $\phi_2=3.49\pm0.29$.
From this we find the following combinations:
$R_{21}=A_2/A_1=0.219\pm0.065$ and $\phi_{21}=\phi_2-2\phi_1=5.99\pm0.31$.
In the case of $n=3$, we would obtain: $A_0=19.591\pm0.009$,
$A_1=0.249\pm0.016$, $A_2=0.053\pm0.016$, $A_3=0.043\pm0.016$,
$\phi_1=5.03\pm0.06$, $\phi_2=3.59\pm0.31$, $\phi_3=5.09\pm0.36$, and hence:
$R_{21}=A_2/A_1=0.213\pm0.066$, $\phi_{21}=\phi_2-2\phi_1=6.09\pm0.33$,
$R_{31}=A_3/A_1=0.173\pm0.065$, $\phi_{31}=\phi_3-3\phi_1=2.56\pm0.41$.
All above uncertainties were calculated from the standard propagation
formula for independent variables. The combination $\phi_{21}$ is the
best parameter to distinguish between the fundamental-mode and
first-overtone pulsators with periods around 5.7~d (log$P \approx 0.76$).
The value of $\phi_{21}$, here very similar for $n=2$ and $n=3$, is evidently
too high for fundamental-mode Cepheids at this period (see Fig.~\ref{fig:phi21}),
which light curves have characteristic saw-tooth shapes
\citep{2008AcA....58..163S,2010AcA....60...17S}. To support our claim,
in Fig.~\ref{fig:steepness}, we present how fast high-amplitude classical
Cepheids increase their brightness from minimum to maximum. For stars with
periods between 2 and 7~d it takes about 20\% of the period, in contrast
to 40\% in S3.

The close-to-sinusoidal light curve shape of S3, or the $\phi_{21}$ value
of $6.0\pm0.3$, could indicate a first-overtone Cepheid,
but $I$-band amplitudes as high as 0.5~mag are not observed at long periods
in this type of pulsators (see Fig.~\ref{fig:pamp}). Moreover, longest observed
periods in first-overtone Cepheids tend to shorten with metallicity.
For instance, periods in this type of Cepheids from the Large Magellanic Cloud
are not longer than 5.9~d \citep{2008AcA....58..163S}, while from more
metal-poor Small Magellanic Cloud are not even longer than 4.5~d
\citep{2010AcA....60...17S}. In none of known environments, $I$-band
amplitudes of the first-overtone Cepheids with periods $>2$~d exceed
0.35~mag. A dwarf galaxy is expected to host old and metal-poor stars,
such as RR Lyrae type variables and anomalous Cepheids rather than
classical ones.

On the other hand, amplitudes of 0.5~mag or higher are observed
in numerous chromospherically active stars, such as of RS CVn type.
In Fig.~\ref{fig:spotted}, we show three examples
of spotted stars with periods of a few days found in the OGLE database.
Light curves of such stars change slowly over years and very often may
mimick pulsations and eclipses if observed for a short period of time.
Based on the observed properties of the light curve, we conclude that
object S3 is very likely a spotted star with the period of 5.695~d,
but the period of 11.390~d cannot be excluded. See comparison of the
Fourier combinations for S3 and spotted stars in Fig.~\ref{fig:S3spotted}.

Object S4, which is too faint to be detected in the deep OGLE GVS images,
according to \cite{2015ApJ...802L...4C} has a period of 13.9~d and a
skewed $K_s$-band light curve with the rising part about twice longer
than the fading one. Such a shape is not observed in classical
Cepheids. \cite{2005PASP..117..823S} normalized near-IR photometry
for 30 Galactic and 31 LMC fundamental-mode Cepheids with periods
mostly longer than 10~d and showed that their $K$-band light curves
are very homogeneous and nearly symmetric. \cite{2012ApJ...748..107P}
performed a global model for 287 classical Cepheids with $P>10$~d
based on thousands of multiband observations. According to their
work $K$-band light curves of stars with periods of 13--14~d have
the rise and fall time scales equal. $K$-band light curves of
long-period classical Cepheids presented in \cite{2004AJ....128.2239P},
\cite{2011ApJS..193...12M}, and \cite{2015AJ....149..117M} are
practically symmetric and none of them shows the rising branch twice
longer then the fading one. Some of the presented light curves have
a small number of points in $K$ and their scatter is comparable with
the amplitude, like OGLE-LMC-CEP-2562 in \cite{2015AJ....149..117M}
and S4 here.

%%%%%%%%%%%%%%%%%%%%%%%%%%%%%%%%%%%%%%%%%%%%%%%%%%%%%%%%%%%%%%%%%%%%%%%%%%%%%%%

\section{Conclusions}

Our observations show that none of the three VISTA objects detected in the
OGLE GVS data is a classical Cepheid. Objects S1 and S2 do not look
to be variable at all. Object S3 is indeed variable, but it has a nearly
sinusoidal light curve with a high amplitude of 0.5~mag in the $I$ band,
likely not stable in time. This is not observed in classical Cepheids
in any environment. S3 is very likely a spotted star of
RS CVn type. Although we are not able to detect star S4,
its $K_s$-band light curve with $P=13.9$~d and the twice longer
rising part than the fading one would be very unusual for a
fundamental-mode classical Cepheid. Based on the above findings we
conclude that there is no evidence for the presence of a dwarf galaxy
behind the Galactic disk in the observed direction.
Our results show that a great care must be taken when classifying
variable objects based on low-amplitude near-infrared light curves,
especially when composed of a small number of data points.

%%%%%%%%%%%%%%%%%%%%%%%%%%%%%%%%%%%%%%%%%%%%%%%%%%%%%%%%%%%%%%%%%%%%%%%%%%%%%%%

\acknowledgements

The OGLE project has received funding from the National Science
Centre, Poland, grant MAESTRO 2014/14/A/ST9/00121 to AU.
This work has been also supported by the Polish Ministry of Sciences
and Higher Education grants No. IP2012 005672 under the Iuventus Plus
program to PP and No. IdP2012 000162 under the Ideas Plus program to IS.

%%%%%%%%%%%%%%%%%%%%%%%%%%%%%%%%%%%%%%%%%%%%%%%%%%%%%%%%%%%%%%%%%%%%%%%%%%%%%%%


\begin{thebibliography}{}

\bibitem[Cardelli et al.(1989)]{1989ApJ...345..245C} Cardelli, J.~A., Clayton, G.~C., \& Mathis, J.~S. 1989, \apj, 345, 245
\bibitem[Chakrabarti et al.(2015)]{2015ApJ...802L...4C} Chakrabarti, S., Saito, R., Quillen, A., et al. 2015, \apj, 802, L4
\bibitem[Feast et al.(2014)]{2014Natur.509..342F} Feast, M.~W., Menzies, J.~W., Matsunaga, N., \& Whitelock, P.~A. 2014, \nat, 509, 342
\bibitem[Macri et al.(2015)]{2015AJ....149..117M} Macri, L.~M., Ngeow, C.-C., Kanbur, S.~M., Mahzooni, S., \& Smitka, M.~T. 2015, \aj, 149, 117
\bibitem[Monson \& Pierce(2011)]{2011ApJS..193...12M} Monson, A.~J., \& Pierce, M.~J. 2011, \apjs, 193, 12
\bibitem[Minniti et al.(2010)]{2010NewA...15..433M} Minniti D., Lucas, P.~W., Emerson, J. et al. 2010, New Astronomy, 15, 433
\bibitem[Pejcha \& Kochanek(2012)]{2012ApJ...748..107P} Pejcha, O., \& Kochanek, C.~S. 2012, \apj, 748, 107
\bibitem[Persson et al.(2004)]{2004AJ....128.2239P} Persson, S.~E., Madore, B.~F., Krzemi\'nski, W., Freedman, W.~L., Roth, M., \& Murphy, D.~C. 2004, \aj, 128, 2239
\bibitem[Pietrukowicz et al.(2013a)]{2013AcA....63..115P} Pietrukowicz, P., Mr\'oz, P., Soszy\'nski, et al. 2013a, Acta Astron., 63, 115
\bibitem[Pietrukowicz et al.(2013b)]{2013AcA....63..379P} Pietrukowicz, P., Dziembowski, W.~A., Mr\'oz, P., et al. 2013b, Acta Astron., 63, 379
\bibitem[Pojma\'nski(2002)]{2002AcA....52..397P} Pojma\'nski, G. 2002, Acta Astron., 52, 397
\bibitem[Schwarzenberg-Czerny(1996)]{1996ApJ...460L.107S} Schwarzenberg-Czerny, A. 1996, \apj, 460, L107
\bibitem[Sitek \& Pojma\'nski(2014)]{2014AcA....64..115S} Sitek, M., \& Pojma\'nski, G. 2014, Acta Astron., 64, 115
\bibitem[Soszy\'nski et al.(2005)]{2005PASP..117..823S} Soszy\'nski, I., Gieren, W., \& Pietrzy\'nski, G. 2005, \pasp, 117, 823
\bibitem[Soszy\'nski et al.(2008)]{2008AcA....58..163S} Soszy\'nski, I., Poleski, R., Udalski, A., et al. 2008, Acta Astron., 58, 163
\bibitem[Soszy\'nski et al.(2010)]{2010AcA....60...17S} Soszy\'nski, I., Poleski, R., Udalski, A., et al. 2010, Acta Astron., 60, 17
\bibitem[Udalski et al.(2015)]{2015AcA....65....1U} Udalski, A., Szyma\'nski, M., \& Szyma\'nski, G. 2015, Acta Astron., 65, 1
\bibitem[Wo\'zniak(2000)]{2000AcA....50..421W} Wo\'zniak, P.~R. 2000, Acta Astron., 50, 421

\end{thebibliography}
\end{document}